\documentstyle[sprocl]{article}
\newcommand{\newc}{\newcommand}

\newc{\gsim}{\lower.7ex\hbox{$\;\stackrel{\textstyle>}{\sim}\;$}}
\newc{\lsim}{\lower.7ex\hbox{$\;\stackrel{\textstyle<}{\sim}\;$}}

\def\beq{\begin{equation}}
\def\eeq{\end{equation}}
\def\beqn{\begin{eqnarray}}
\def\eeqn{\end{eqnarray}}

\def\inbar{\,\vrule height1.5ex width.4pt depth0pt}

\def\IC{\relax\hbox{$\inbar\kern-.3em{\rm C}$}}
\def\IQ{\relax\hbox{$\inbar\kern-.3em{\rm Q}$}}
\def\IR{\relax{\rm I\kern-.18em R}}
 \font\cmss=cmss10 \font\cmsss=cmss10 at 7pt
\def\IZ{\relax\ifmmode\mathchoice
 {\hbox{\cmss Z\kern-.4em Z}}{\hbox{\cmss Z\kern-.4em Z}}
 {\lower.9pt\hbox{\cmsss Z\kern-.4em Z}}
 {\lower1.2pt\hbox{\cmsss Z\kern-.4em Z}}\else{\cmss Z\kern-.4em Z}\fi}
\def\NPB#1#2#3{{\it Nucl.\ Phys.}\/ {\bf B#1} (19#2) #3}
\def\PLB#1#2#3{{\it Phys.\ Lett.}\/ {\bf B#1} (19#2) #3}
\def\PRD#1#2#3{{\it Phys.\ Rev.}\/ {\bf D#1} (19#2) #3}

\def\beq{\begin{equation}}
\def\eeq{\end{equation}}
\def\beqn{\begin{eqnarray}}
\def\eeqn{\end{eqnarray}}
\def\ie{{\it i.e.}\/}
\def\eg{{\it e.g.}\/}
\begin{document}
\rightline{CERN-TH/98-245}
\rightline{{\tt hep-ph/9807522}}
\rightline{July 1998}
\bigskip
\title{TeV-Scale GUTs~\footnote{Invited talks given by
     KRD at PASCOS '98
     (Boston, USA, 23--29 March 1998);
      by TG at the European Meeting ``From the Planck
      Scale to the Electroweak Scale'' (Kazimierz, Poland, 24--30 May 1998);
      and by KRD and TG at SUSY '98
      (Oxford, England, 11--17 July 1998).}}
\author{\sc Keith R. Dienes$^{1}$, $\,$
       Emilian Dudas$^{1,2}$, $\,$and$\,$ Tony Gherghetta$^{1}$ }
\address{ ~\\
    $^1$ CERN Theory Division, CH-1211 Geneva 23, Switzerland\\
    $^2$ LPTHE, Univ.\ Paris-Sud, F-91405 Orsay Cedex, France }
\maketitle\abstracts{
    In this talk, we summarize our recent proposal for
    lowering the scale of grand unification to the TeV range
    though the appearance of extra spacetime dimensions.
    Particular emphasis is placed on the perturbativity
    and predictivity of our scenario, as well
    as its sensitivity to unification-scale effects.
    }
\input epsf.tex

\section{Introduction}

The possibility of extra large spacetime dimensions has
recently received considerable
attention~\cite{antoniadis,Dim,shortpaper,Dimtwo,henry,longpaper,Dimthree,bachas}.
One of the first serious investigations of the phenomenological
properties of theories with extra large dimensions was
made in Ref.~\cite{antoniadis}, where the lightness
of the supersymmetry-breaking scale was related to the largeness
of a string-theoretic compactification radius.
More recently, in Ref.~\cite{Dim}, extra large dimensions figured
prominently in a radical
new proposal for avoiding the gauge hierarchy problem by lowering
the Planck scale to the TeV scale.
The extra dimensions required to achieve this are in the
(sub-)millimeter
range, and imply a profound
change in Newton's gravitational force law
at such distances.  Likewise, in
Refs.~\cite{Dimtwo,henry,longpaper,bachas}, it was shown how extra
large dimensions could also be used to lower the fundamental string
scale to the TeV scale.  The idea of taking the string scale
in the TeV range originates in Ref.~\cite{lykken}, and makes
use of special features of open-string theory first pointed
out in Ref.~\cite{Witten}.

Besides the Planck and string scales, there also exists one
additional high fundamental scale in physics:  the GUT scale.
Indeed, while the Planck and (perturbative heterotic) string scales
are related directly to each other, the GUT scale stands independently.
In this talk, therefore, we will concentrate on our complementary
proposal~\cite{shortpaper,longpaper}
of lowering the fundamental GUT scale to the TeV range.
By definition,
one of the unique issues faced in attempting to lower the GUT scale
to the TeV range is that the scale of gauge coupling
unification must be substantially
shifted from its usual MSSM value near $M_{\rm GUT}\approx 2\times 10^{16}$
GeV.
This in turn requires that the usual logarithmic running of the MSSM
gauge couplings must somehow be altered.  Remarkably, however, as
demonstrated in Refs.~\cite{shortpaper,longpaper}, extra
spacetime dimensions
have precisely the effect we want:  they modify the running of the
gauge couplings in such a way that not only is the unification preserved,
but in fact it occurs more rapidly.
This then leads to a reduction in the unification scale.
Thus,
through extra dimensions,
it becomes possible for the first time to contemplate grand unification
occurring at an intermediate scale --- indeed, as low as the TeV scale.
In other words, we now see that
it becomes possible to replace a four-dimensional GUT
at $10^{16}$ GeV with a higher-dimensional GUT at the TeV scale.

There is one aspect of this proposal that deserves special attention.
By its very nature, such a radical change in the GUT scale requires
that the running of the gauge couplings fully experience the effects
of the extra dimensions.  This in turn requires that the Standard
Model fields actually propagate in the extra dimensions,
and moreover that the effects of these extra dimensions not be arranged
to cancel as the result of the effects of
other symmetries.
In this respect, the extra dimensions required in this
proposal are fundamentally different
from those of Ref.~\cite{antoniadis}
(in which the effects of the extra dimensions on the gauge couplings
are arranged to cancel as the result of $N=4$ supersymmetry);
from those of Ref.~\cite{Dim}
(with respect to which the Standard Model particles are trapped on
an effective brane);  and also from those of Ref.~\cite{bachas} (in which the
effects of the
extra dimensions are restricted to certain open-string sectors).
As far as we are aware, therefore,
the effects of extra dimensions on the running of the gauge couplings had
not been investigated prior to Refs.~\cite{shortpaper,longpaper}, and
as we shall see, this leads to special subtleties that we shall discuss below.

\section{Gauge Coupling Unification in the Presence of Extra Dimensions}

We begin by quickly recalling the usual four-dimensional result.
In the usual MSSM calculation, one derives the one-loop
running of the $SU(3)\times SU(2)\times U(1)$ gauge couplings
by evaluating one-loop wavefunction renormalization diagrams,
where all MSSM states can propagate in the loop.
This then leads to the logarithmic running equation
\beq
         \alpha_i^{-1}(\mu) ~=~
         \alpha_i^{-1}(M_Z) ~-~ {b_i\over 2\pi}\,\ln {\mu\over M_Z}~.
\label{oneloopRGEfour}
\eeq
where the one-loop MSSM beta-function
coefficients are $(b_1,b_2,b_3)= (33/5,1,-3)$.
Given the experimentally measured gauge couplings
at the $Z$-scale, we can use Eq.~(\ref{oneloopRGEfour}) to extrapolate
upwards in energy.  This then leads to the conventional unification
near $M_{\rm GUT}\approx 2\times 10^{16}$ GeV.

How do we extend this calculation into higher dimensions?
The basic idea is relatively simple.
First, we imagine that there exist $\delta\equiv D-4$
extra spacetime dimensions, each compactified on a circle of
radius $R$.  Here $\delta=1,2,...$ can
take any integer value, and likewise $\mu_0\equiv R^{-1}$, which sets
the energy threshold for the extra dimensions, can range
anywhere from several hundred GeV all the way up to the usual GUT
scale $M_{\rm GUT}\approx 2\times 10^{16}$ GeV.
Corresponding to each MSSM field, there will be an infinite
tower of Kaluza-Klein states whose masses are separated by $\mu_0$.
For various technical reasons discussed in
Refs.~\cite{shortpaper,longpaper}, we actually must compactify
on orbifolds rather than circles, and likewise there are some
subtleties involved in arranging the Kaluza-Klein towers corresponding
to the MSSM states.  The upshot is that the states at the excited levels
of the Kaluza-Klein towers must fall into $N=2$ supermultiplets (even though
the ground-state zero-mode MSSM states are only $N=1$ supersymmetric).
It also turns out that one has the freedom
to choose whether or not to have Kaluza-Klein towers for the MSSM
fermions.  We shall therefore denote by the additional free parameter
$\eta=0,1,2,3$ the number of MSSM chiral generations which we shall
assume to have Kaluza-Klein towers.

Given this setup, it is then relatively straightforward to calculate
the running of the gauge couplings:  we simply re-evaluate the
standard one-loop wavefunction renormalization diagrams
with the usual MSSM states as well as their
corresponding Kaluza-Klein excitations
in the loop.   Of course, strictly speaking,
such a theory is non-renormalizable due to the infinite towers of Kaluza-Klein
states, but we are free to truncate these towers at some arbitrarily high
excitation level without affecting our results.  Therefore, for the purposes
of calculating gauge coupling renormalization effects, we may consider
this to be a fully renormalizable field theory.
Evaluating the one-loop diagrams, we then find the general
result~\cite{longpaper}
\beq
       \alpha_i^{-1}(\Lambda) ~=~ \alpha_i^{-1}(\mu) ~-~
            {b_i-\tilde b_i\over 2\pi}\,\ln{\Lambda\over \mu}
          ~-~ {\tilde b_i\over 4\pi}\,
             \int_{r\Lambda^{-2}}^{r\mu^{-2}} {dt\over t} \,
     \left\lbrace \vartheta_3\left( {it\over \pi R^2} \right)
                                       \right\rbrace^\delta~
\label{KKresult}
\eeq
where the Jacobi theta-function
$\vartheta_3(\tau) \equiv \sum_{n_i=-\infty}^\infty e^{\pi i \tau  n^2}$
reflects the sum over the Kaluza-Klein states.
Here the new beta-function coefficients $\tilde b_i$ corresponding to the
excited Kaluza-Klein levels are given by
\beq
               (\tilde b_1,\tilde b_2,\tilde b_3) ~=~ (3/5, -3,-6)~
           + ~\eta\,(4,4,4)~,
\eeq
and the numerical coefficient $r$ is given by $r\equiv \pi
(X_\delta)^{-2/\delta}$
where $X_\delta\equiv 2\pi^{\delta/2}/\delta\Gamma(\delta/2)$.
While the final term in Eq.~(\ref{KKresult}) reflects the contributions
from the Kaluza-Klein towers, the penultimate term reflects the fact
that the zero-mode MSSM states and their Kaluza-Klein towers are not
identical.~\cite{longpaper}   Note that
all of the terms in Eq.~(\ref{KKresult}) together form the one-loop
result, and thus this expression is
renormalization-group scheme-independent.

For many practical purposes, it is possible to {\it approximate}\/ the
result (\ref{KKresult}) in the following way:  for $\mu\leq \mu_0$, we may
replace Eq.~(\ref{KKresult}) by Eq.~(\ref{oneloopRGEfour}), while
for $\mu\geq \mu_0$, we may replace Eq.~(\ref{KKresult})
by the power-law expression
\beq
     \alpha_i^{-1}(\mu) ~=~
      \alpha_i^{-1}(\mu_0)
      ~-~ {b_i-\tilde b_i \over 2\pi} \,\ln{\mu\over\mu_0}
           ~-~ {\tilde b_i X_\delta\over 2\pi \delta} \,\left\lbrack
                 \left({\mu\over \mu_0}\right)^\delta -1\right\rbrack~.
\label{geffd}
\eeq
However, for certain precision calculations
(as will be discussed below), it may be necessary
to use the full one-loop result given in Eq.~(\ref{KKresult}).

\begin{figure}[ht]
\centerline{ \epsfxsize 3.0 truein \epsfbox {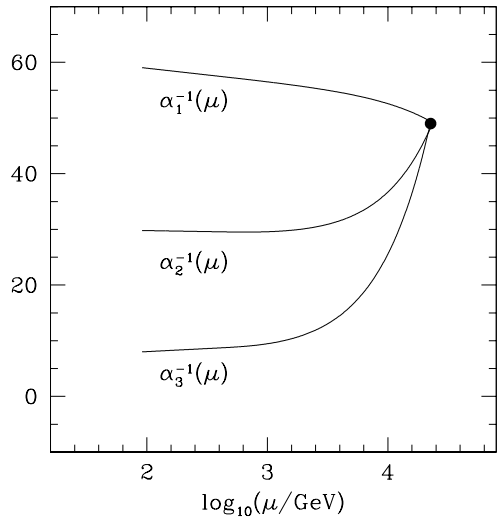}}
\caption{Unification of gauge couplings at
     the new unification scale $M'_{\rm GUT}\approx 20$ TeV,
    assuming the appearance of a single extra spacetime dimension
    of radius $R^{-1}=1$ TeV.}
\label{figureone}
\end{figure}

The remarkable feature of this higher-dimensional running is that
gauge coupling unification is nevertheless preserved.
As the most interesting case, let us consider $\mu_0 = 1$ TeV,
$\delta=1$, and $\eta=0$.  We then find the unification shown
in Fig.~\ref{figureone}.
Thus, we see that extra dimensions are consistent with the emergence
of a grand unified theory in the TeV range --- \ie, the GUT scale
has been lowered all the way to the TeV scale!
However, this is not the only possibility, and it is shown
in Refs.~\cite{shortpaper,longpaper} that such a unification with
a reduced unification scale occurs regardless of the values
of $\mu_0\equiv R^{-1}$, $\delta$, or $\eta$.

\section{Perturbativity, Higher-Loop Corrections, and Sensitivity to\\
      Unification-Scale Thresholds}

Before proceeding further, it is important to discuss this unification
in some detail.
Specifically, the power-law running raises a number
of important questions.
Why does the unification occur?
How robust is it against higher-order corrections?
How perturbative is the resulting theory?
How exact is the unification?
How predictive is it?
And most importantly, how sensitive is it to unification-scale effects?

First, we must discuss {\it why}\/ the unification occurs.
It is clear that if it had been the case that $b_i=\tilde b_i$ for all $i$,
then unification would have occurred as a direct consequence of the usual
unification in the MSSM, for the appearance of the extra Kaluza-Klein towers
with beta-functions $\tilde b_i$ would have simply resembled additional copies
of the MSSM matter content.  In our case,
by contrast, we actually have $b_i\not= \tilde b_i$.
However, continued unification does not require $b_i = \tilde b_i$.  Instead,
all that is required is that $B_{ij}\equiv (\tilde b_i-\tilde b_j)/(b_i-b_j)$
be independent if $(i,j)$, or equivalently
that $B_{12}/B_{13}=1$ and $B_{13}/B_{23}=1$.
We find that in our scenario, these relations hold {\it approximately}\/:
$B_{12}/B_{13}\approx 0.94$ and $B_{13}/B_{23}\approx 0.92$.
Indeed, it is apparent from Fig.~\ref{figureone} that
this leads to a fairly precise unification, and increasing the inverse
radius $R^{-1}$
only makes the unification more precise.
It is this numerical coincidence which underlies our unification.

Given that this (approximate) one-loop unification occurs, it is natural to ask
about higher-loop corrections.  {\it A priori}\/, they might be expected
to be particularly substantial in our case, since they would also seem to have
power-law behavior.  However, in our scenario, the excited Kaluza-Klein
states are always in $N=2$ supermultiplets.  This means that all higher-order
power-law contributions vanish identically, which in turn implies that
one-loop power-law
behavior is exact.  Thus, all that remains are the usual MSSM higher-order
logarithmic effects, which are strongly suppressed.

\begin{figure}[ht]
\centerline{ \epsfxsize 3.0 truein \epsfbox {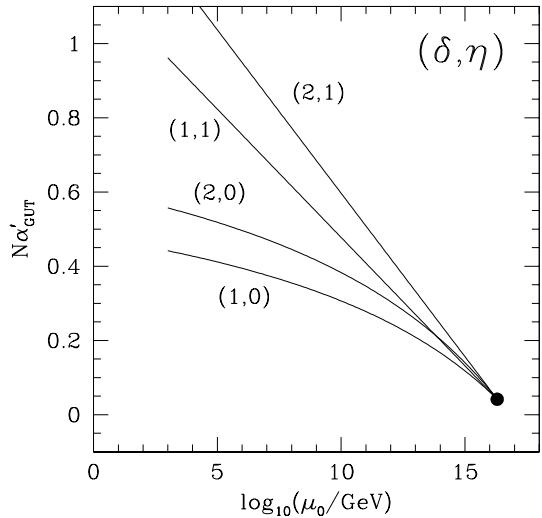}}
\caption{Values of $N\alpha'_{\rm GUT}$
       for different values of $(\delta,\eta)$,
    as a function of $\mu_0\equiv R^{-1}$.
     In all cases, the perturbativity constraint
         $N\alpha'_{\rm GUT}\ll 4\pi$ is easily satisfied.}
\label{figuretwo}
\end{figure}

Next, let us discuss the perturbativity of our scenario.
It is apparent from Fig.~\ref{figureone}
that the unification is
 very {\it weakly coupled}\/ (even more so than within the MSSM).
This is true even if we change the values of $R$ and $\delta$.
However, strictly speaking, this does not imply that the unification is
 {\it perturbative}\/.
Indeed, even though our new unified coupling $\alpha'_{\rm GUT}$ satisfies
$\alpha'_{\rm GUT}\ll 4\pi$, the perturbation-theory expansion
parameter is really $N\alpha'_{\rm GUT}$ where $N\sim (M'_{\rm
GUT}/\mu_0)^\delta$
is the number of Kaluza-Klein states propagating in the loops.
Thus, for true perturbativity,
we must really require $N\alpha'_{\rm GUT}\ll 4\pi$.
However, in Fig.~\ref{figuretwo}, we show the value of
$N\alpha'_{\rm GUT}$ that arises in our scenario for different values
of radii $\mu_0$, $\delta$, and $\eta$.
We see that the constraint
$N\alpha'_{\rm GUT}\ll 4\pi$ is easily satisfied.
Thus, our unification is not only weakly coupled, but also perturbative.

Finally, let us now discuss issues pertaining to the {\it precision}\/ of our
unification.
We have already remarked above that the unification is only
approximate because the Kaluza-Klein beta-functions $\tilde b_i$ do
not have differences whose ratios exactly match those of the MSSM
beta-functions.
How serious a problem is this?

One immediate observation suggests that this problem is severe.
Let us suppose that we run $\alpha_1$ and $\alpha_2$ up to their unification
point,
demand exact unification with $\alpha_3$ at that point,
and then run $\alpha_3$ back down to the $Z$-scale.  Carrying
out this calculation using the {\it full}\/ result in Eq.~(\ref{KKresult}),
we find $\alpha_3(M_Z)\approx 0.174$, which is many standard deviations
away from the experimental result.  However, in our scenario, it does not
take a significant threshold effect at the unification scale to remedy the
situation.  For example, it is easy to show that a mere 6\% threshold
effect at the unification scale can eliminate this
discrepancy.  Such threshold
effects can easily arise from SUSY thresholds,
GUT thresholds, MSSM higher-order logarithmic corrections,
and holomorphic anomaly contributions.
They can also arise
in string theory due to the contributions of heavy string states.
Of course, these corrections can be calculated only when
a complete theory at $M'_{\rm GUT}$ is specified.

However, this then raises a new question.
Because of the steepness of the slopes of the curves in Fig.~\ref{figureone}
near the unification point, it is natural to worry that our scenario
might be unreasonably {\it sensitive}\/ to unification-scale effects.
For example, if we wish to ensure that $\alpha_3(M_Z)$ remains within
its experimental uncertainties, how fine-tuned must these threshold effects
be?  Clearly a fine-tuning factor of $10^{-6}$ or more would be unacceptable.
However, once again, it is straightforward to show that no major fine-tuning
is involved.  Specifically, in order to quantify any potential fine-tuning
parameter that might be involved in our scenario due to the steepness of
the curves in Fig.~\ref{figureone}, we can consider how much of a shift
$\Delta M$ in the unification scale can be tolerated if $\alpha_3(M_Z)$ is
to remain within acceptable limits.  Taking
$\Delta \alpha_3(M_Z)\approx 0.003$ as the experimental uncertainty
around the central value, we find a corresponding unification-scale
sensitivity $\Delta M\approx 0.13$ TeV for $\mu_0= 1$ TeV.
This amounts to a fine-tuning of the order of one part in ten.

Thus, we conclude that our unification scenario is predictive, perturbative,
and not unreasonably sensitive to unification-scale effects.

\section{New Questions Posed by TeV-Scale GUTs}

Given this TeV-scale unification, a number of questions immediately arise.
We shall not have space here to discuss these questions in detail, so we
shall merely summarize some of the results that can be found in
Refs.~\cite{shortpaper,longpaper}.

\subsection{Proton Decay}

First, if we contemplate the appearance of a TeV-scale GUT, we immediately
face the problem of proton-decay mediated by light $X$ and $Y$ bosons and
Higgs(ino) triplets.  However, these contributions can be
cancelled~\cite{longpaper}
to all orders in perturbation theory as a result of the symmetry
properties of the higher dimensions (essentially a parity argument for
the fifth dimension).  This is then an intrinsically higher-dimensional
solution to the proton-decay problem.

\subsection{Fermion Masses and  Soft Masses}

Just like the gauge couplings, the Yukawa couplings and soft masses
will now also experience power-law behavior.  Can this be
used to explain, for example, the fermion mass hierarchy?
Of course, extra dimensions are universal, and will
not introduce a flavor dependence by themselves.   Thus, some
flavor dependence still must be introduced (\eg, in the unification-scale
theory).  But the important point is that this flavor dependence
need not be large, because the power-law running due to the extra
spacetime dimensions can {\it amplify}\/ the effects of even small
flavor-dependent couplings.  This is discussed in more
detail in Ref.~\cite{longpaper}.

\subsection{Need for Supersymmetry?}

Given that new dimensions could appear at a TeV (thereby eliminating
the obvious gauge hierarchy problem),
one might also question the need for supersymmetry.  Can the gauge couplings
also unify at the TeV scale without supersymmetry?  It is found that this can
indeed occur in some circumstances~\cite{longpaper}.   This then suggests
the possibility of {\it non}\/-supersymmetric TeV-scale GUTs.

\subsection{Embeddings into String Theory}

Given gauge coupling unification in the TeV range, it is natural to consider
embedding this scenario into a string theory whose fundamental string scale is
also reduced to the TeV scale.  For technical reasons, such a string would
have to be an open string.\footnote{
     In Ref.~\cite{bachas}, it is pointed out that reduced-scale open
     strings do not necessarily imply power-law running for the
     gauge couplings.  However, our point is the converse:
     power-law running for the gauge couplings --- which is
     the only way to lower the GUT scale --- requires reduced-scale open
     string theories, and is in fact the generic situation for
     such strings.}
The possibility of such TeV-scale open strings
is discussed in Refs.~\cite{lykken,Dimtwo,henry,longpaper}, and a
specific embedding of our scenario into TeV-scale strings is discussed
in Ref.~\cite{longpaper}.

\subsection{Next Directions}

Clearly, this is only the tip of the iceberg.
On the theoretical side, many aspects of physics beyond the Standard Model
must now be considered in a new light.
These include issues pertaining to supersymmetry, supersymmetry-breaking,
GUT physics, and string theory.  For each, we must now investigate the
role of extra large dimensions, and the effects of the drastically altered
energy scales.
Even more excitingly, on the experimental side,
many striking signals (such as the production, detection, decays,
and indirect effects of TeV-scale Kaluza-Klein states)
would be observable at future colliders.  Many of these
signals are discussed in Ref.~\cite{antoniadis}.
Thus, it becomes possible to consider probing the properties of GUTs
and strings {\it experimentally}\/, thereby giving rise to a possible new
 {\it experimental}\/ direction for string phenomenology!
Finally, because the fundamental energy scales
of physics have changed, there will also be profound effects
for cosmology.

As we have said, this is only the tip of the iceberg.
However, as is evident from our results as well
as those of Refs.~\cite{Dim,Dimtwo},
the important point is that extra spacetime dimensions provide a natural
way of bringing fundamental physics down to low (and perhaps
even accessible) energy scales.
This titanic shift in thinking will have many consequences as yet
unseen.  Let us hope that our ship can explore this vast iceberg
without sinking.


\section*{Acknowledgments}

We are happy to thank
S.~Abel, I.~Antoniadis, R.~Barbieri,
P.~Bin\'etruy, S.~Dimopoulos, M.~Dine,
A.~Ghinculov, G.~Kane, S.~King, C.~Kounnas, J.~Lykken,
J.~March-Russell, J.~Pati, M.~Quir\'os,
S.~Raby, L.~Randall, R.~Rattazzi, G.~Ross,
S.-H.H.~Tye, and C.~Wagner
for questions and comments on our work.


\end{document}